\documentclass[%
 aip,
 amsmath,amssymb,
 reprint,%
]{revtex4-1}

\usepackage{graphicx}
\usepackage{dcolumn}
\usepackage{bm}

\usepackage[utf8]{inputenc}
\usepackage[T1]{fontenc}
\usepackage{mathptmx}
\usepackage{etoolbox}
\usepackage{xcolor}

\makeatletter
\def\@email#1#2{%
 \endgroup
 \patchcmd{\titleblock@produce}
  {\frontmatter@RRAPformat}
  {\frontmatter@RRAPformat{\produce@RRAP{*#1\href{mailto:#2}{#2}}}\frontmatter@RRAPformat}
  {}{}
}%
\makeatother
\begin{document}

\preprint{AIP/123-QED}

\title{A moment tensor potential for lattice thermal conductivity \\ calculations of $\alpha$ and $\beta$ phases of Ga$_{2}$O$_{3}$}
\author{Nikita~Rybin}
 \email{N.Rybin@skoltech.ru}
 \affiliation{Skolkovo Institute of Science and Technology,\\ Bolshoy Boulevard 30, bld. 1, 121205, Moscow, Russia}
\author{Alexander~Shapeev}
 \email{A.Shapeev@skoltech.ru}
\affiliation{Skolkovo Institute of Science and Technology,\\ Bolshoy Boulevard 30, bld. 1, 121205, Moscow, Russia}%

\date{\today}

\begin{abstract}
Calculations of heat transport in crystalline materials have recently become mainstream, thanks to machine-learned interatomic potentials that allow for significant computational cost reductions while maintaining the accuracy of first-principles calculations. Moment tensor potentials (MTP) are among the most efficient and accurate models in this regard. In this study, we demonstrate the application of MTP to the calculation of the lattice thermal conductivity of $\alpha$ and $\beta$-Ga$_{2}$O$_{3}$.  Although MTP is commonly employed for lattice thermal conductivity calculations, the advantages of applying the active learning methodology for potential generation is often overlooked. Here, we emphasize its importance and illustrate how it enables the generation of a robust and accurate interatomic potential while maintaining a moderate-sized training dataset. 
\end{abstract}

\maketitle

\section{Introduction}

Lattice thermal conductivity (LTC) is an essential quantity for a wide range of technological applications, including thermal management of microelectronic devices~\cite{Moore2014}, thermoelectrics~\cite{Tan2016}, and thermal barrier coating materials~\cite{Clarke2005}. Accurate experimental measurement of LTC is challenging and reproducibility between different experimental groups is often not guaranteed~\cite{Wei2016}. Consequently, despite the technological need, reliable LTC data are known for only a small number of materials. At the same time, the theoretical formulation of heat transport constitutes a mature research field~\cite{Peierls1929, Maradudin1962}. From a practical perspective, calculations of the temperature-dependent LTC are usually based on a perturbative approach, the Green-Kubo approach (GK)~\cite{Kubo1957, Meng2019}, or non-equilibrium molecular dynamics simulations (NEMD). In a perturbative approach~\cite{Broido2007}, calculations are typically performed within the framework of first-principles (or \textit{ab initio}) density-functional theory (DFT).  In practice, perturbative approaches require calculation of the so-called interatomic force constants (IFCs), which is time-consuming, especially for structures with large unit cells and low symmetry. This is because the number of structures with displaced atoms needed to construct third-order IFCs grows enormously fast with reduction of the crystalline symmetry~\cite{Togo2023}.

Materials with large anharmonicities are not rare and sometimes accurate evaluation of lattice dynamics of such materials require to consider the inclusion of all orders of anharmonicity~\cite{Knoop2020}. Already the influence of four-phonon scattering processes on LTC of solids can have huge impact~\cite{Feng2016, Feng2017}. For example, in the case of boron arsenide (BAs) and graphene, better agreement between theory and experiment can only be achieved when four-phonon scattering is taken into account~\cite{Kang2018, Tian2018, Li2018, Wang2023}. In principle, lattice dynamics anharmonicity can be fully and accurately taken into account in \textit{ab initio} molecular dynamics (AIMD) simulations. Hence, both the GK and NEMD approaches utilize AIMD and allow for a more accurate computation of LTC, as all orders of anharmonicity are taken into account. However, these methods are computationally expensive. The Green-Kubo formula relates instantaneous fluctuations of heat current in terms of the autocorrelation function. While the formulation of GK was done long ago, first-principles calculations of LTC using this approach have only recently appeared~\cite{Carbogno2017}. The evaluation of LTC using the Green-Kubo approach coupled with first-principles calculations is rare, albeit its profound influence on the understanding of the heat transport~\cite{Knoop2023}. However, converging LTC values is a challenging task, as AIMD simulations are severely limited by system size (a few hundred atoms) and timescales (tens of picoseconds). Thus, GK and NEMD approaches are limited in applicability when coupled with first-principles calculations. 

Challenges associated with the experimental measurements and theoretical calculations of the LTC lead to the lack of a systematic understanding of the LTC beyond semi-empirical and phenomenological trends in a very limited number of simple materials~\cite{Morelli2006}. From the computational perspective, speed up can be obtained by performing molecular dynamics (MD) simulations using empirical potentials, which enables one to efficiently predict properties of both bulk materials and nanostructures, over a wider range of system sizes (up to hundreds of nanometers)~\cite{Kadau2006} and timescales (up to tens of nanoseconds)~\cite{Bockmann2002}. However, the prediction accuracy will be highly dependent on the fidelity of the potentials. On the other hand, machine learning methods (ML) can be used to predict LTC values. In some cases ML models are trained in a supervised manner~\cite{Jaafreh2021, Purcell2023, Luo2023} and in some cases machine-learned interatomic potentials (MLIPs)~\cite{Deringer2019, Bartok2017} are used to either compute IFCs~\cite{Mortazavi2020, Mortazavi2021, Ouyang2022} or to directly perform MD simulations~\cite{Mortazavi2020, Arabha2021, Verdi2021, Korotaev2019}. Among different MLIP models the moment tensor potential (MTP)~\cite{Shapeev2016} showed substantial accuracy keeping the required training data at a minimum size~\cite{Zuo2020}, especially once the active learning (sometimes terms active sampling or learning on-the-fly are used) methodology is used~\cite{Podryabinkin2017}. Despite wide application of different MLIPs, the training process is still frequently done without the active learning scheme. In this case the potential is usually trained on some AIMD trajectories and its ability to extrapolate is typically limited only to this training set. Recently, we demonstrated that low root-mean-squared errors on the training and test sets do not guarantee the robustness of a potential~\cite{Rybin2024_flinak}, which we define as an ability to run long MD simulations without failure. This outcome underscores that while low energy and force errors are achieved, applicability and accuracy of such MLIP should be taken with a portion of scepticism, even if the lattice dynamics calculations are performed under the same thermodynamic conditions as those employed during the training data generation. 

In this work, we demonstrate how to generate a robust and accurate MTP using the active learning approach. Its employment not only ensures the applicability of MTP for lattice dynamics simulations, but also substantially reduces the number of the required DFT calculations. As a benchmark systems, we used $\alpha$- and $\beta$-Ga$_{2}$O$_{3}$. Both $\alpha$ and $\beta$ phases of Ga$_{2}$O$_{3}$ are wide-bandgap semiconductors of significant technological importance~\cite{Shan2005, Pearton2018, Higashiwaki2017, Singh2017, Fleischer1992, Zhou2017, Cheng2020}. Consequently, these materials are well-investigated both experimentally~\cite{Guo2015, Galazka2014, Handwerg2015, Jiang2018} and theoretically~\cite{Santia2015, Yan2018}. Moreover, given the large conventional cell of $\beta$-Ga$_{2}$O$_{3}$ containing 20 atoms, the first-principles calculations are highly time-consuming especially when extensive convergence tasks need to be performed, which highlights the role of MLIPs usage as would be demonstrated here.

\begin{figure*}[hb!]
	\centering
	\begin{minipage}[h]{0.455\linewidth}
    \includegraphics[trim={0cm 0cm 0cm 0cm},clip, width=1\linewidth]{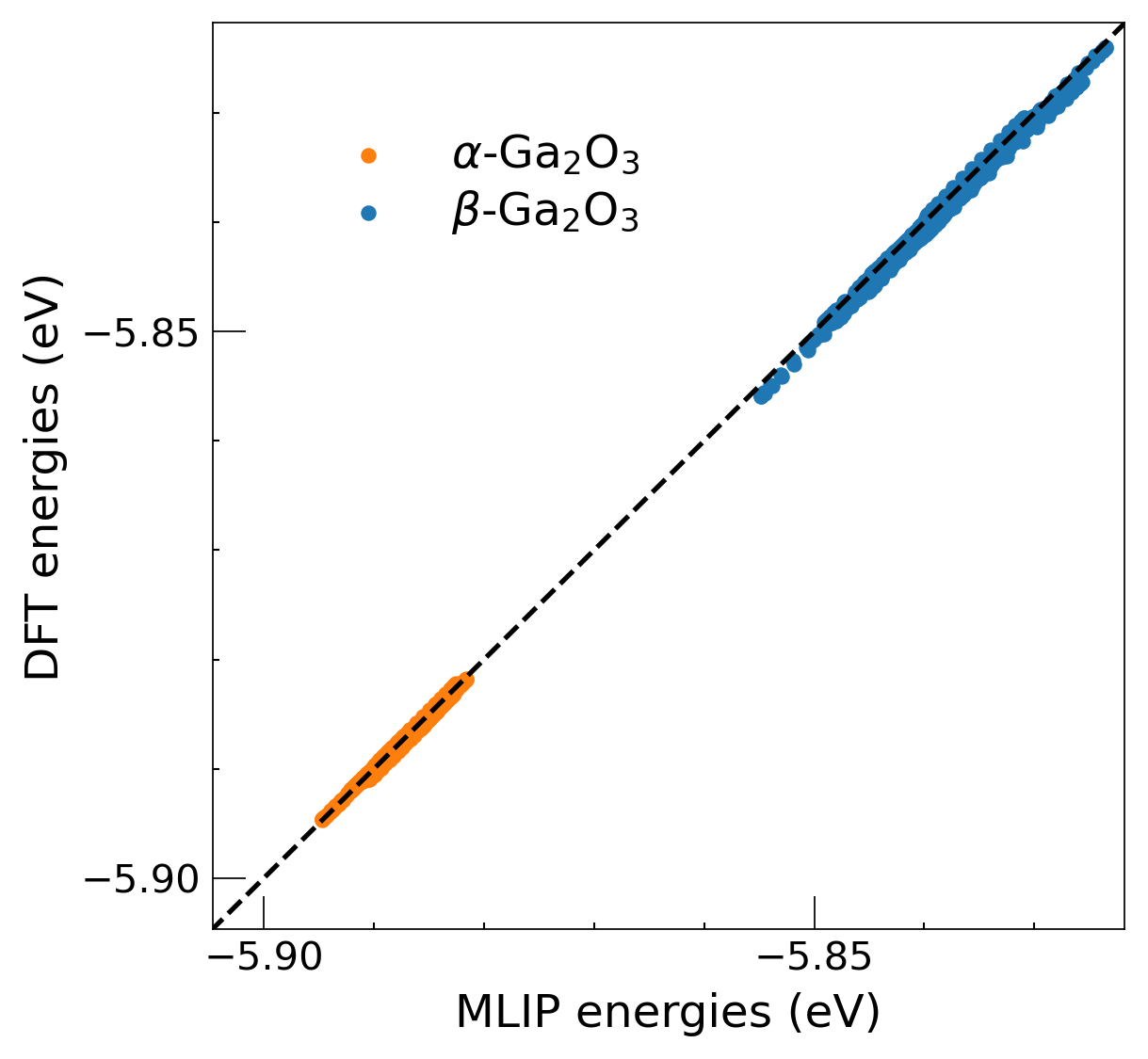} \\ (a)
	\end{minipage}
 	\begin{minipage}[h]{0.43\linewidth}
	\includegraphics[trim={0cm 0cm 0cm 0cm},clip, width=1\linewidth]{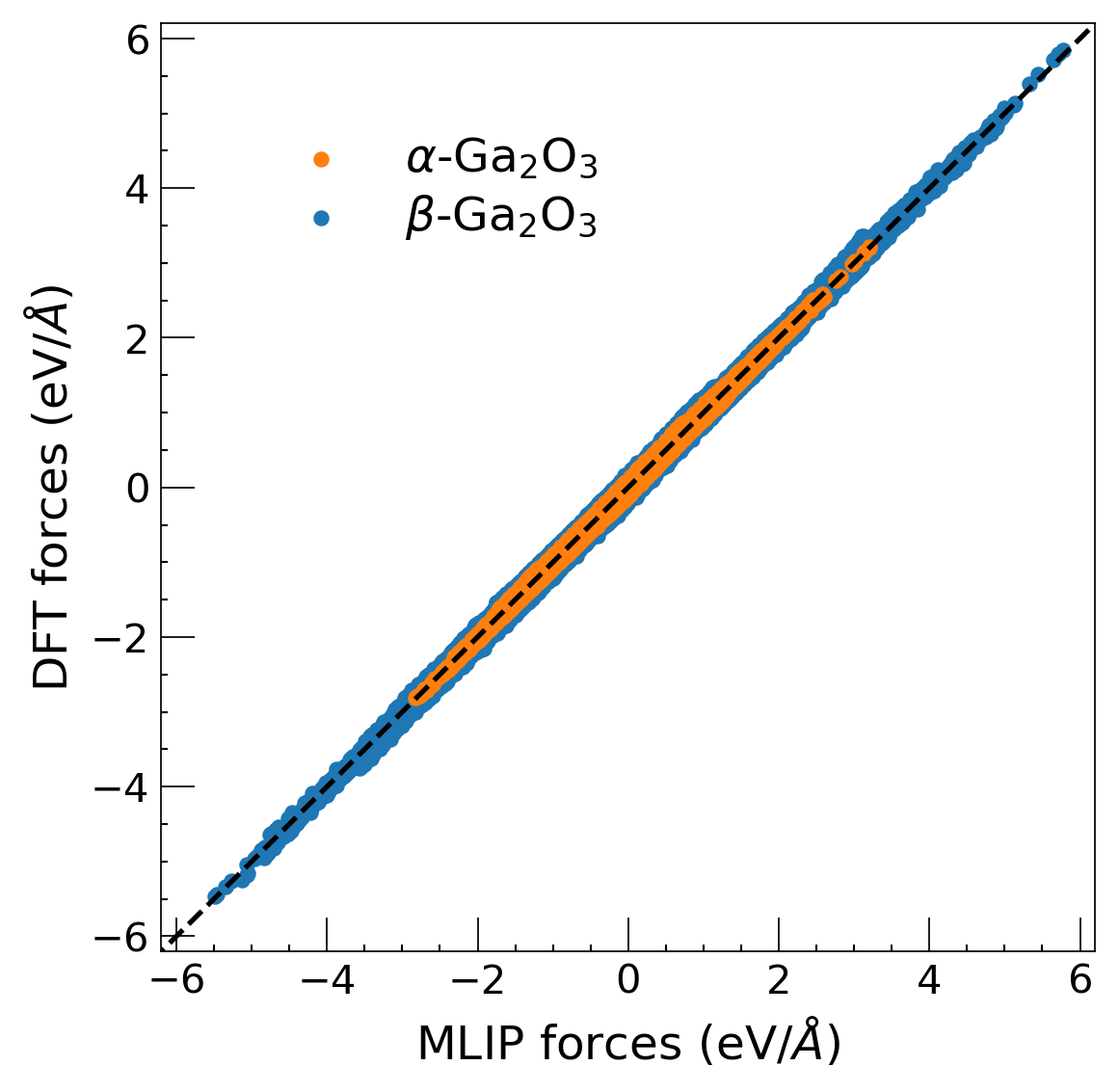}  \\ (b)
	\end{minipage} 
	\caption{Comparison between DFT- and MTP-calculated (a) energies and (b) forces for both $\alpha$- and $\beta$-Ga$_{2}$O$_{3}$ structures. Straight black line represents a perfect linear dependence.}
	\label{fig:en_f_comparison} 
\end{figure*}

\begin{figure*}[hb!]
	\centering
	\begin{minipage}[h]{0.45\linewidth}
    \includegraphics[trim={0cm 0cm 0cm 0cm},clip, width=1\linewidth]{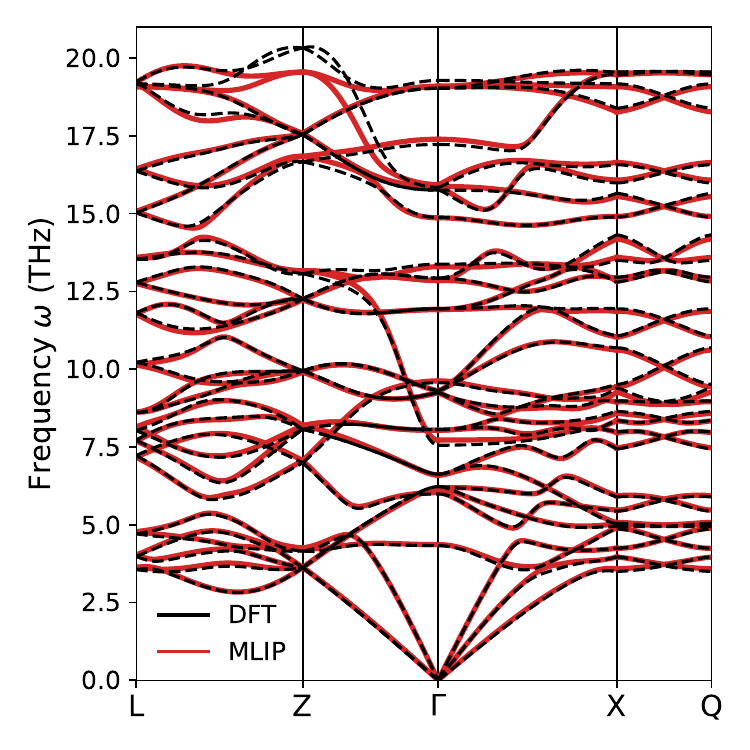} \\ (a)
	\end{minipage}
 	\begin{minipage}[h]{0.45\linewidth}
	\includegraphics[trim={0cm 0cm 0cm 0cm},clip, width=1\linewidth]{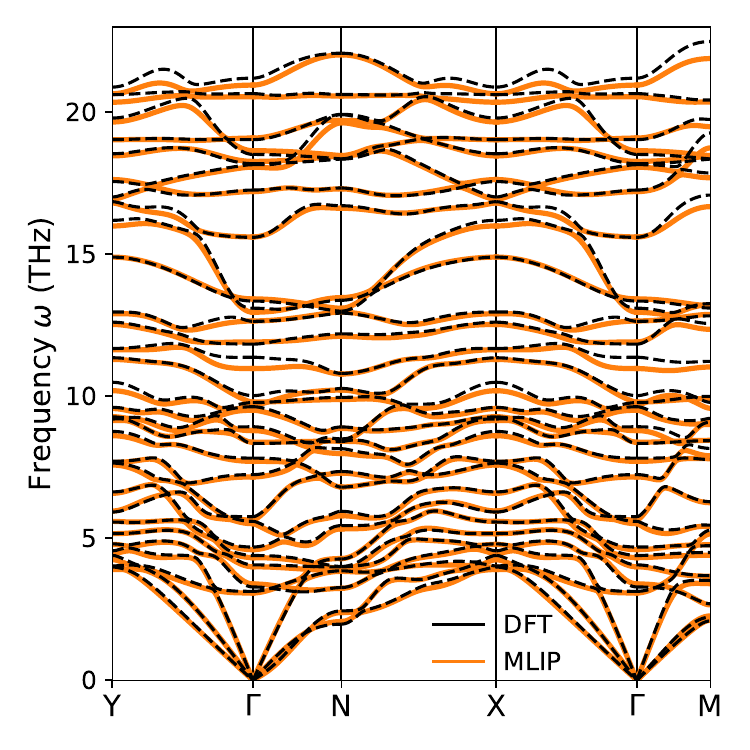} \\ (b)
	\end{minipage} 
	\caption{Phonon dispersion relation of (a) $\alpha$- and (b) $\beta$-Ga$_{2}$O$_{3}$ along selected high symmetry Brillouin zone path. Calculations are done using DFT and MTP.}
	\label{fig:phonons} 
\end{figure*}

\begin{figure*}[hb!]
	\centering
	\begin{minipage}[h]{0.48\linewidth}
		\center{\includegraphics[trim={0cm 0cm 0cm 0cm},clip, width=1\linewidth]{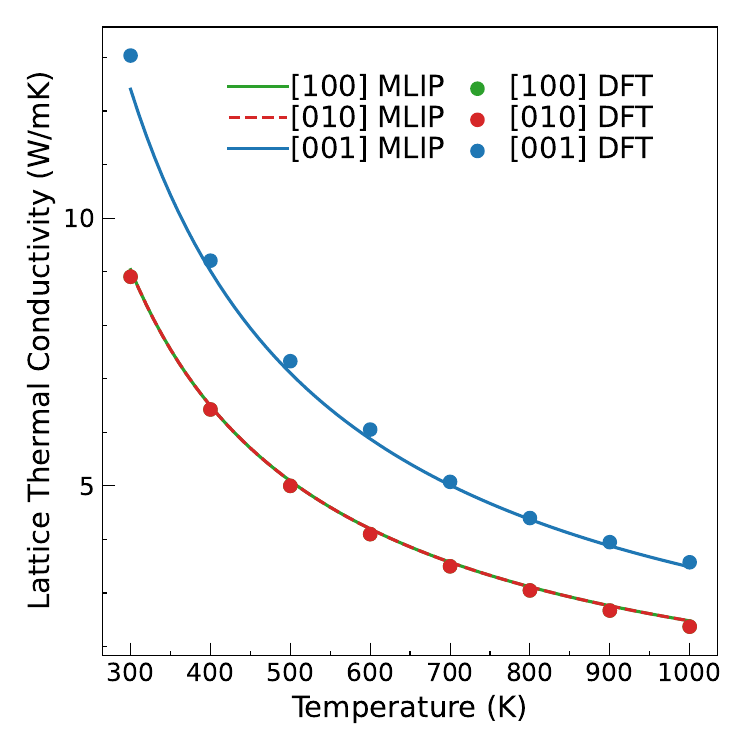}} \\ (a)
	\end{minipage}
 	\begin{minipage}[h]{0.48\linewidth}
		\center{\includegraphics[trim={0cm 0cm 0cm 0cm},clip, width=1\linewidth]{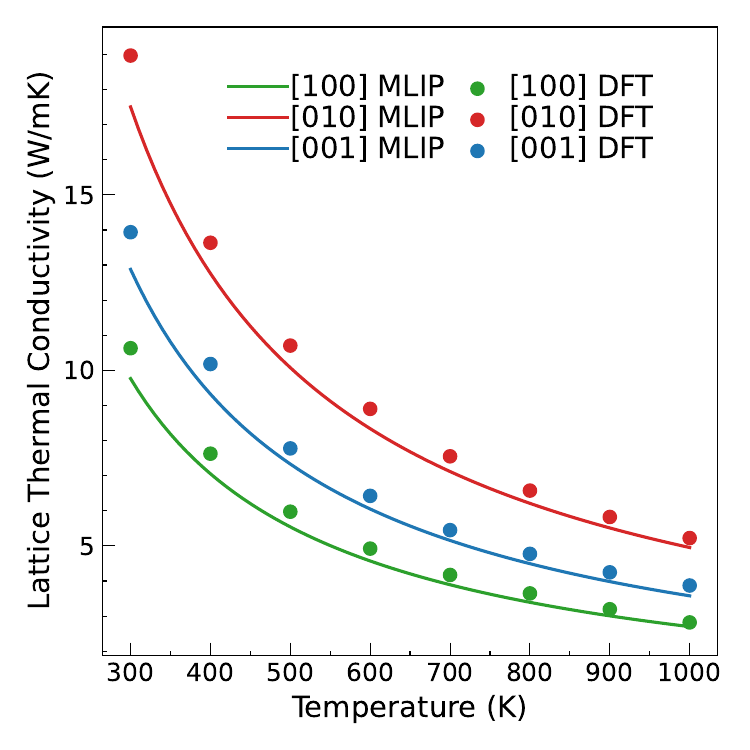}} \\ (b)
	\end{minipage}
	\caption{Temperature dependence of the LTC of (a) $\alpha$-Ga$_{2}$O$_{3}$ and (b) $\beta$-Ga$_{2}$O$_{3}$ for three different directions. LTC calculations with MLIP (moment tensor potential in our case) are compared to the recent calculations, which were done in the DFT framework~\cite{Yang2024}.}
	\label{fig:ltc}
\end{figure*}

\section{Methodology}

\subsection{Construction of a Moment Tensor Potential}

The potential energy of an atomic system as described by the MTP interatomic potential is defined as a sum of the energies of atomic environments of the individual atoms:

\[
    E_{\text{MTP}} = \sum_{i=1}^{N} V(n_{i}),
\]
where the index $i$ label $N$ atoms of the system, and $n_{i}$ describes the local atomic neighborhood around atom \textit{i} within a certain cutoff radius $R_\text{cut}$, i.e., a many-body descriptor can comprise the information of all neighbors of a centered atom up to a given cutoff radius. The function $V$ is the moment tensor potential: 
\[
    V(n_{i}) = \sum_{\alpha} \xi_{\alpha} B_{\alpha}(n_{i}),
\]
where $\xi_{\alpha}$ are the fitting parameters and $B_{\alpha}(n_{i})$ are the basis functions that will be defined below. Moment tensors descriptors are used as representations of atomic environments and defined as: 
\[
    M_{\mu, \nu}\left({n}_i\right)=\sum_j f_\mu\left(\left|r_{i j}\right|, z_i, z_j\right) \underbrace{r_{i j} \otimes \ldots \otimes r_{i j}}_{\nu \text { times }},
\]
where the index $j$ goes through all the neighbors of atom $i$. The symbol ``$\otimes$'' stands for the outer product of vectors, thus ${r}_{i j} \otimes \cdots \otimes {r}_{i j}$ is the tensor of rank $\nu$ encoding the angular part which itself resembles moments of inertia. 
The function $f_\mu$ represents the radial component of the moment tensor:
\[
f_\mu\left(\left|r_{i j}\right|, z_i, z_j\right)=\sum_k c_{\mu, z_i, z_j}^{(k)} Q^{(k)}(r),
\]
where $z_i$ and $z_j$ denote the atomic species of atoms $i$ and $j$, respectively, $r_{ij}$ describes the positioning of atom $j$ relative to atom $i$, $c_{\mu, z_i, z_j}^{(k)}$ are the fitting parameters and 
\[
Q^{(k)}(r):=T_k(r)\left(R_{\text {cut }}-r\right)^2
\]
are the radial functions consisting of the Chebyshev polynomials $T_k(r)$ on the interval $[R_\text{min},  R_\text{cut}]$ with the term ($R_\text{cut} - r)^2$ that is introduced to ensure a smooth cut-off to zero. The descriptors $M_{\mu, \nu}$ taking $\nu$ equal to $0, 1, 2, \ldots$ are tensors of different ranks that allow to define basis functions as all possible contractions of these tensors to a scalar, for instance:
\[
\begin{aligned}
& B_0\left({n}_i\right)=M_{0,0}\left({n}_i\right) \\
& B_1\left({n}_i\right)=M_{0,1}\left({n}_i\right) \cdot M_{0,1}\left({n}_i\right), \\
& B_2\left({n}_i\right)=M_{0,0}\left({n}_i\right)\left(M_{0,2}\left({n}_i\right): M_{0,2}\left({n}_i\right)\right) .
\end{aligned}
\]
Therefore the level of $M_{\mu, \nu}$ is defined by ${\rm lev}M_{\mu, \nu}$ = 2$\mu$ + $\nu$ and if $B_{\alpha}$ is obtained from $M_{\mu_1, \nu_1}$ , $M_{\mu_2, \nu_2}$ ,
$\dots$, then ${\rm lev}B_{\alpha}$ = $(2\mu_1 + \nu_1)$ + $(2\mu_2 + \nu_2)$ + $\dots$ . By including all basis functions such that ${\rm lev}B_{\alpha} < d$ we obtain the moment tensor potential of level $d$ which we denote as MTP$_d$.

As fundamental symmetry requirements, all descriptors for atomic environment have to be invariant to translation, rotation, and permutation with respect to the atomic indexing. Particularly for structurally intricate Ga$_{2}$O$_{3}$ polymorphs, the incorporation of many-body interactions is crucial to ensuring the accuracy of the derived potentials. The MTP framework satisfies all aforementioned conditions~\cite{Shapeev2016}.  Importantly, the MTP framework satisfies all aforementioned conditions~\cite{Shapeev2016}. MTP methodology and tools to utilize it are implemented in the MLIP-2 package~\cite{Novikov2021}, which we used in this work.

\subsection{Generation of a Training Set}

We conducted aiMD simulations to collect the reference data (potential energy, atomic forces, atomic coordinates)) for MTP training. aiMD simulations are conducted within the DFT framework, employing VASP (Vienna \textit{ab initio} simulations package)~\cite{Kresse1996} with the projector augmented wave method~\cite{Kresse1999}. The Perdew-Burke-Ernzerhof generalized gradient approximation (PBE-GGA)~\cite{Perdew1996} was employed for the exchange–correlation functional. We used a plane wave basis set cut-off of 550~eV, and a Gaussian smearing of 0.1~eV width. The halting criterion for electronic density convergence during self-consistent field calculations is set to 10$^{-7}$~eV. 

The corundum structure of $\alpha$-Ga$_{2}$O$_{3}$ has a space group $R\bar{3}c$, while $\beta$-Ga$_{2}$O$_{3}$ has a monoclinic structure with the space group $C2/m$. The optimized structural parameters obtained in our calculations are shown in the Tab.~\ref{tab:ga2o3_structure}. These structural parameters are in a good agreement with the experimental data~\cite{Ahman1996}. All lattice dynamics calculations in this work were performed with these structures. The conventional cells was sampled using 6$\times$6$\times$2 k-point grids for $\alpha$-Ga$_{2}$O$_{3}$ and for $\beta$-Ga$_{2}$O$_{3}$, respectively. In case of the lattice dynamics calculations, we used 2$\times$2$\times$1 diagonal transformation matrix to generate a supercell with 120 atoms from the conventional $\alpha$-Ga$_{2}$O$_{3}$ cell. In the case of $\beta$-Ga$_{2}$O$_{3}$ we generated a supercell with 160 atoms expanding the conventional cell using 6$\times$6$\times$2 diagonal transformation matrix. For the calculations with supercells, the k-point grids were adjusted to keep the k-points' mesh density as in the calculations with the unit cells, this leads to the utilization of only one $\Gamma$-point in both cases.

\begin{table}[]
\caption{PBE-optimized unit cell vectors and angles of the $\alpha$- and $\beta$-Ga$_{2}$O$_{3}$ phases.}
\begin{tabular}{ccccccc}
\hline
\multicolumn{1}{c|}{Phase} & a, (\AA)                    & b, (\AA)                  & c, (\AA)                      & $\alpha$                & $\beta$                 & $\gamma$                \\ \hline
\multicolumn{1}{c|}{$\alpha$-Ga$_{2}$O$_{3}$} & 5.065                & 5.065                & 13.64                & 90$^{\circ}$                   & 90$^{\circ}$                   & 120$^{\circ}$                  \\
\multicolumn{1}{c|}{$\beta$-Ga$_{2}$O$_{3}$}  & 12.469$^{\circ}$               & 3.087                & 5.882                & 90$^{\circ}$                   & 103.673$^{\circ}$              & 90$^{\circ}$                   \\ \hline
\multicolumn{1}{l}{}       & \multicolumn{1}{l}{} & \multicolumn{1}{l}{} & \multicolumn{1}{l}{} & \multicolumn{1}{l}{} & \multicolumn{1}{l}{} & \multicolumn{1}{l}{}
\end{tabular}
\label{tab:ga2o3_structure}
\end{table}

Typically, two main approaches are employed to train a reliable MLIP capable of accurately representing a wide spectrum of local atomic environments. The first one involves the inclusion of various systems, necessitating extensive AIMD simulations. By incorporating a diverse range of systems, the MLIP can effectively learn to capture the nuances of different atomic environments. The second approach is called active learning and it involves the selective addition of data points to the training set, focusing on structures where the MLIP extrapolates significantly. These are instances where the MLIP's predictions for energies and forces exhibit high uncertainty. By strategically choosing a minimal yet diverse set of training data in the feature space, this active learning strategy enables effective fitting of the potential and helps overcome extrapolation challenges. In this study, we generate a small dataset to pretrain the MLIP. Subsequently, we employ an active learning strategy to iteratively expand this dataset, ensuring the MLIP's robustness and accuracy for lattice dynamics calculations. We advocate for the consistent use of the active learning methodology due to its efficacy in enhancing the MLIP's performance and thus, revisit it briefly here. 

Various active learning schemes are employed in the development of machine learning potentials~\cite{Zhang2019, Sivaraman2020}. In our study, we utilized the D-optimality-based active learning procedure developed in~\cite{Podryabinkin2017} and available in the MLIP-2 package. The initial training set was generated in aiMD simulations with isothermal-isobaric (NPT) ensemble using the Nos\'{e}-Hoover thermostat~\cite{Nose1984}. The initial dataset was generated at T=700~K for $\alpha$ phase and at T=1800~K for $\beta$ phase, respectively. Each AIMD trajectory was simulated for 2~ps with a 1~fs time step. Initial structures for AIMD were prethermalized following the algorithm established in~\cite{West2006}, which allows to equilibrate simulations at finite temperatures faster. The first picosecond of each AIMD simulation was discarded, and the second picosecond was subsampled with the 5~fs time intervals, resulting in a set of 200 snapshots for each structure. These samples were employed to train an initial MTP of level 12. 

Within the active learning algorithm, we initiate MD calculations in LAMMPS (Large-scale Atomic/Molecular Massively Parallel Simulator)~\cite{Thompson2022} using a pretrained MTP as the model for interatomic interactions in the NVE ensemble. At each step of the MD simulations, the algorithm assesses the extrapolation grade $\gamma$ of the atomic configuration. The D-optimality criterion guides the decision to include a configuration in the training set, based solely on atomic coordinates. This unique feature, coupled with the linear form of the potential, allows MTP to learn effectively on-the-fly. Configurations with $\gamma > 2$ are added to the preselected set. When $\gamma$ exceeds 10, the MD simulation halts, and all sufficiently different configurations from the preselected set are incorporated into the training, followed by the refitting of the potential. This procedure repeats until MD simulations can run without failure for 30~ps. In our case, MTP achieved robustness with a training set comprising 439 samples in the case of $\alpha$-Ga$_{2}$O$_{3}$ and 403 samples in the case of $\beta$-Ga$_{2}$O$_{3}$. This amount of samples is two times to an order of magnitude smaller than for other machine-learning potentials~\cite{Li2018,Liu2020}. The potential trained in this manner demonstrated the ability to robustly conduct MD simulations for 100~ps, showcasing the effectiveness of the active learning procedure in terms of reducing the necessary DFT data for training and enhancing potential robustness. The training set generated during the active learning procedure was then used to train MTP with level 16. 

The fitting accuracy of obtained potential can be verified by the agreement between MTP predictions and the benchmark AIMD results. We compare the MTP-predicted energies and forces for a separate testing dataset formed by running AIMD for 1~ps with time steps of 1~fs (this dataset contains 1000 snapshots) at T=700~K for $\alpha$ and T=1800~K for $\beta$ phases, respectively. It is found that our MTP can provide the root mean square errors of 0.2 and 0.6 meV/atom for the potential energy of $\alpha$ and $\beta$ phases. The RMSEs for the atomic forces are 22 meV/\AA\ and 46 meV/\AA\ for $\alpha$ and $\beta$ phases, respectively. Therefore, MTP can accurately reproduce $\textit{ab initio}$ energies and forces as also visually demonstrated in Fig.~\ref{fig:en_f_comparison}(a,~b).

\section{Results and Discussions}

Harmonic or second-order force constants can be used to evaluate the accuracy of MTP compared to to DFT. We thus calculate the phonon dispersion relation along selected high symmetry paths, using the supercell approach and the finite displacement method~\cite{Parlinski1997}, as implemented in the Phonopy package~\cite{Togo2015}. Figs.~\ref{fig:phonons}~(a,~b) show calculated phonon dispersion relations obtained using MTP and DFT for both polymorphs of Ga$_{2}$O$_{3}$, respectively. We note good agreement between both methods, which validates an ability of MTP to describe the lattice dynamics in the harmonic approximation. Notably, existing empirical potentials frequently fail to capture that~\cite{Blanco2005, Ma2020}. The most observable differences in the phonon spectrum might be seen for high-frequency optical phonon modes (especially at the Z-point in Fig.~\ref{fig:phonons}~(a)). This is a usual problem frequently arising in the harmonic constants simulations with MLIPs~\cite{Mortazavi2020, Li2018}. From our experience, it is possible to improve the accuracy of optical phonon modes evaluation by performing simulations at high temperatures for a set of compressed and expanded structures~\cite{Mortazavi2021, Liu2020}. In this work we skipped this procedure, since obtaining precise values of LTC is not the primarily goal of this work. Considering the low symmetry and high anisotropy of $\beta$-Ga$_{2}$O$_{3}$, it is reasonable to see that the phonon dispersion profile is much more complex than that of many other wide-bandgap materials, such as GaN and ZnO~\cite{Wu2016}. Due to the versatility stemming from the relatively high complexity of the MTP functional form, harmonic force constants can be accurately reproduced --- this precision is crucial as it serves as a fundamental prerequisite for conducting calculations of the LTC.

We next proceed towards LTC calculations, which was evaluated by solving Boltzmann transport equation for phonons in the relaxation time approximation as implemented in the Phono3py package~\cite{Togo2015_2, Togo2018, Togo2023}. Here we should mention that in the case of $\alpha$-Ga$_{2}$O$_{3}$ one has to compute forces acting on atoms in 4087 supercells (with 120 atoms) and in the case of $\beta$-Ga$_{2}$O$_{3}$ in 15225 supercells (with 160 atoms). Albeit the crystalline symmetry is taken into account to reduce the number of calculations, these numbers are tangible even for calculations on modern high-performance clusters. Performing these calculations with MTP as a model for interatomic interactions takes minutes on a personal workstation. 

Figs.~\ref{fig:ltc}~(a,~b) present temperature dependence of the LTC for $\alpha$ and $\beta$ polymorphs. Values of LTC obtained using MTP are in a good agreement with most recent DFT calculations~\cite{Yang2024}, for example, at 300~K LTC values along [100] are identical up to the second decimal. Since $\beta$-Ga$_{2}$O$_{3}$ has a monoclinic structure, its [001] direction does not align with the $z$-axis. Therefore, we calculated its LTC in the [001] direction as was done in previous works~\cite{Li2018, Yang2024}: $\kappa_{[001]}$=$\cos^{2}{\beta}\kappa_{xx} + \sin{2\beta}\kappa_{xz}+\sin^{2}{\beta}\kappa_{zz}$. Notably, MTP caught the anisotropy in LTC values along different directions in both $\alpha$- and $\beta$-Ga$_{2}$O$_{3}$, which further validates its accuracy. The difference between our results and previous theoretical calculations~\cite{Yang2024} is within 5-10\%, which might be easily associated with different structural parameters or calculations set up between studies. Taken into account tremendous reduction of simulation time, we do not consider this as a significant problem.

\section{Conclusion}

In summary, we developed MTPs for atomistic simulations of $\alpha$ and $\beta$-Ga$_{2}$O$_{3}$. Our potentials exhibit good accuracy in reproducing the \textit{ab initio} PES, attaining a total energy accuracy below 1~meV/atom and force accuracy below 50~meV/\AA\ for both polymorphs. We then compared phonon dispersion and LTC values calculated using MTP with the results of DFT, and obtained good agreement, which highlights the applicability of MTP in calculating the lattice dynamics and heat transport of solids. The proper choice of training data to generate an accurate potential using as little data as possible is governed by the utilization of the active learning methodology. Although the possibility of its usage is frequently ignored, the active learning scheme plays a crucial role in enhancing the robustness of the machine-learned interatomic potentials, as demonstrated here.  

\begin{acknowledgments}
We acknowledge funding from the Russian Science Foundation (Project No. 23-13-00332).
\end{acknowledgments}

\section*{Data Availability Statement}
The data that support the findings of this study are available from the corresponding author upon reasonable request.

\clearpage
\bibliography{aipsamp}

\end{document}